\documentclass[%
 reprint,
unsortedaddress,
 amsmath,amssymb,
 aps,
]{revtex4-1}

\usepackage{graphicx}% Include figure files
\usepackage{dcolumn}% Align table columns on decimal point
\usepackage{bm}% bold math
\usepackage{siunitx}
\usepackage[version=4]{mhchem}
\usepackage[%
colorlinks,bookmarksopen,bookmarksnumbered,citecolor=blue,urlcolor=blue,anchorcolor=blue,linkcolor=blue
]{hyperref}
\usepackage{amsmath}
\usepackage{xcolor}
\begin{document}

%\preprint{APS/123-QED}

\title{Tuning Rashba spin-orbit coupling at \ce{LaAlO3}/\ce{SrTiO3} interfaces by band filling}% Force line breaks with \\

\author{Chunhai Yin$ ^{1} $, Patrick Seiler$ ^{2} $, Lucas M. K. Tang$ ^{3} $, Inge Leermakers$ ^{3} $, Nikita Lebedev$ ^{1} $, Uli Zeitler$ ^{3} $, and Jan Aarts$ ^{1} $}
\affiliation{
	$ ^{1} $Huygens-Kamerlingh Onnes Laboratory, Leiden University, P.O. Box 9504, 2300 RA Leiden, The Netherlands\\
	$ ^{2} $Center for Electronic Correlations and Magnetism, EP VI, Insitute of Physics, University of Augsburg, 86135 Augsburg, Germany\\
	$ ^{3} $High Field Magnet Laboratory (HFML-EMFL), Radboud University, Toernooiveld 7, 6525 ED Nijmegen, The Netherlands\\
}

\begin{abstract}
The electric-field tunable Rashba spin-orbit coupling at the \ce{LaAlO3}/\ce{SrTiO3} interface shows potential applications in spintronic devices. However, different gate dependence of the coupling strength has been reported in experiments. On the theoretical side, it has been predicted that the largest Rashba effect appears at the crossing point of the $d_{xy}$ and $d_{xz,yz}$ bands. In this work, we study the tunability of the Rashba effect in \ce{LaAlO3}/\ce{SrTiO3} by means of back-gating. The Lifshitz transition was crossed multiple times by tuning the gate voltage so that the Fermi energy is tuned to approach or depart from the band crossing. By analyzing the weak antilocalization behavior in the magnetoresistance, we find that the maximum spin-orbit coupling effect occurs when the Fermi energy is near the Lifshitz point. Moreover, we find strong evidence for a single spin winding at the Fermi surface. 
\end{abstract}

\maketitle

Complex oxide heterostructures provide an interesting platform for novel physics since their physical properties are determined by correlated $d$ electrons \cite{hwang2012nm}. The most famous example is the discovery of a high mobility two-dimensional electron system (2DES) at the interface between \ce{LaAlO3} (LAO) and \ce{SrTiO3} (STO) \cite{ohtomo2004}. Intriguing properties, such as superconductivity \cite{reyren2007}, signatures of magnetism \cite{brinkman2007nm,lee2013NM} and even their coexistence \cite{li2011,bert2011}, have been reported.

At the LAO/STO interface, the 2DES is confined in an asymmetric quantum well (QW) in STO. The intrinsic structure inversion asymmetry introduces an electric field which gives rise to a Rashba spin-orbit (SO) coupling \cite{bychkov1984}. Additionally, due to the large dielectric constant of the STO substrate at cryogenic temperatures \cite{neville1972JAP}, the coupling constant can be tuned with the STO as back gate \cite{shalom2010PRL, caviglia2010PRL, liang2015PRB}. This could give rise to applications in spintronics, such as spin field-effect transistors \cite{datta1990APL}. However, the reported results are inconsistent. Upon increasing the back-gate voltage ($V\rm_{G}$), the SO coupling strength was found to decrease \cite{shalom2010PRL}, increase \cite{caviglia2010PRL}, or show a maximum \cite{liang2015PRB}. A clear understanding of the SO coupling dependence on $V\rm_{G}$ is necessary for more advanced experiments.

For a free electron gas the Rashba spin splitting is proportional to the symmetry breaking electric field. However the Rashba effect in solids like semiconductor and oxide heterostructures has a more complicated origin \cite{winkler2003}. Theoretical studies have shown that multi-band effects play an essential role in the SO coupling in LAO/STO \cite{zhong2013PRB,khalsa2013PRB}. At the LAO/STO (001) interface, the band structure is formed by the \ce{Ti} $t_{2g}$ bands. At the $\Gamma$-point, the $d_{xy}$ band lies below the $d_{xz,yz}$ bands in energy \cite{santander2011nature}. Applying $V\rm_{G}$ across the STO substrate changes the carrier density and therefore the Fermi energy ($E\rm_{F}$). A Lifshitz transition occurs when $E\rm_{F}$ is tuned across the bottom of the $d_{xz,yz}$ bands \cite{joshua2012NC}. The largest SO coupling effect was predicted at the crossing point of the $d_{xy}$ and $d_{xz,yz}$ bands \cite{zhong2013PRB,khalsa2013PRB}. The SO coupling theory was experimentally confirmed later by angle-resolved photoemission spectroscopy (ARPES) measurements \cite{king2014NC}. 

So far, few experiments actually track the evolution of SO coupling when $E\rm_{F}$ is driven to approach or depart from the Lifshitz point. In this work, we study the Rashba effect in back-gated LAO/STO. By carefully monitoring the sign of the magnetoresistance (MR) in high magnetic field and the linearity of the Hall resistance, $V\rm_{G}$ was tuned back and forth so that the Lifshitz transition was crossed multiple times. The SO coupling characteristic magnetic fields were extracted by fitting the weak antilocalization (WAL) behavior in the MR. We find that the maximum SO coupling effect occurs when $E\rm_{F}$ is near the Lifshitz point. We also find a single spin winding at the Fermi surface.

\begin{figure}[t]
	\centering
	\includegraphics[width=0.96\linewidth]{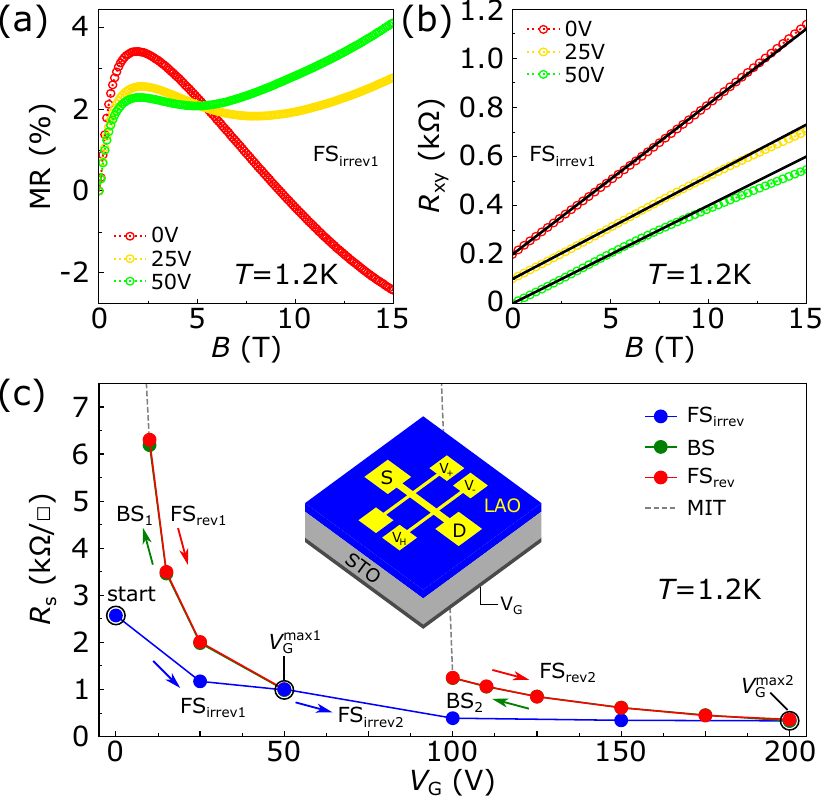}
	\caption{(a) Magnetoresistance (MR) and (b) Hall resistance ($R\rm_{xy}$) as a function of magnetic field ($B$) in the first irreversible forward sweep (FS$\rm_{irrev1}$) at \SI{1.2}{\kelvin}. $R\rm_{xy}$($B$) curves are separated by an offset and the black lines are linear fits to them. (c) Sheet resistance ($R\rm_{s}$) as a function of $V\rm_{G}$ at \SI{1.2}{\kelvin}. FS$\rm_{irrev}$, BS, and FS$\rm_{rev}$ stand for irreversible forward sweep, backward sweep and reversible forward sweep, respectively. Two BSs were performed at \SI{50}{\volt} ($V\rm_{G}^{max1}$) and \SI{200}{\volt} ($V\rm_{G}^{max2}$). Note that BS and FS$\rm_{rev}$ overlap perfectly. Inset shows a schematic of the Hall bar device. Source and drain are labeled as S and D. The longitudinal resistance ($R\rm_{xx}$) is measured between V$\rm_{+}$ and V$\rm_{-}$ and the transverse resistance ($R\rm_{xy}$) between V$\rm_{H}$ and V$\rm_{-}$. $V\rm_{G}$ is applied between the back of the STO substrate and the drain.}
	\label{figRV}
\end{figure}

We use a Hall bar device with a width of $W=$ \SI{150}{\micro\meter} and length of $L=$ \SI{1000}{\micro\meter}, as depicted in the inset of Fig. \ref{figRV}(c). First, a sputtered amorphous AlO$_{x}$ hard mask in form of a negative Hall bar geometry (thickness $\sim$\SI{15}{\nano\meter}) was fabricated on a \ce{TiO2}-terminated STO (001) substrate by photolithography. Then, 15 unit cells of LAO film were deposited at \SI{800}{\degreeCelsius} in an Ar pressure of \SI{0.04}{\milli\bar} by \SI{90}{\degree} off-axis sputtering \cite{yin2019prm}. Finally, the sample was $in\,situ$ annealed at \SI{600}{\degreeCelsius} in \SI{1}{\milli\bar} of oxygen for \SI{1}{\hour}. The back-gate electrode was formed by uniformly applying a thin layer of silver paint (Ted Pella, Inc.) on the back of the substrate. The detailed device fabrication procedure is described in Ref. \cite{2019trap}.
Magnetotransport measurements were performed in a cryostat with a base temperature of \SI{1.2}{\kelvin} and a magnetic field of \SI{15}{\tesla}. The longitudinal resistance ($R\rm_{xx}$) and transverse resistance ($R\rm_{xy}$) were measured simultaneously using standard lock-in technique ($f=$ \SI{13.53}{\hertz} and $i\rm_{RMS}=$ \SI{1.0}{\micro\ampere}). The maximum applied $V\rm_{G}$ was \SI{200}{\volt} and the leakage current was less than \SI{1.0}{\nano\ampere} during the measurement. 

The device was first cooled down to \SI{1.2}{\kelvin} with $V\rm_{G}$ grounded. In the original state ($V\rm_{G}$ = \SI{0}{\volt}), the observed maximum in MR (Fig \ref{figRV}(a)) in low magnetic field is a sign of WAL. The negative MR in high magnetic field as well as the approximately linear $R\rm_{xy}$($B$) (Fig \ref{figRV}(b)) indicate that the presence of only one type of carriers. Next, $V\rm_{G}$ was increased to add electrons to the QW and two characteristic Lifshitz transition features appeared at \SI{25}{\volt}. They are the emergence of positive MR in high magnetic field and the change of linearity of $R\rm_{xy}$($B$) \cite{joshua2012NC,smink2017PRL}. $V\rm_{G}$ was further increased to \SI{50}{\volt} ($V\rm_{G}^{max1}$) to drive $E\rm_{F}$ slightly above the Lifshitz point, resulting in larger positive MR and more downward bending of $R\rm_{xy}$($B$) in high magnetic field.

\begin{figure}[t]
	\centering
	\includegraphics[width=0.96\linewidth]{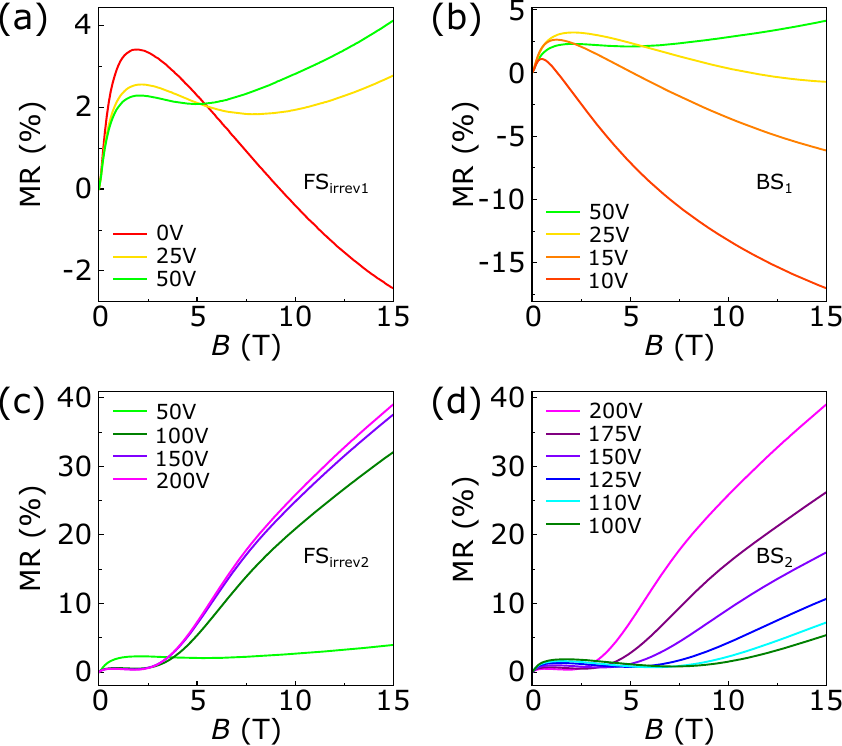}
	\caption{Back-gate tuning of MR in various regimes (a) FS$\rm_{irrev1}$, (b) BS$\rm_{1}$, (c)  FS$\rm_{irrev2}$, and (d) BS$\rm_{2}$. Data for reversible forward sweeps are omitted since they show similar behaviors as backwards sweeps.}
	\label{figmr}
\end{figure}

Then $V\rm_{G}$ was decreased to remove electrons from the QW in order to go back through the Lifshitz transition from the high-density direction. It has been shown that, due to the effect of electron trapping in STO, the sheet resistance ($R\rm_{s}$) always follows an irreversible route when $V\rm_{G}$ is first swept forward and then backward \cite{2019trap,bell2009prl,liu2015APLM}. Fig. \ref{figRV}(c) shows $R\rm_{s}$ as a function of $V\rm_{G}$. It can be seen that $R\rm_{s}$ increases above the virgin curve when $V\rm_{G}$ is swept backward. The backward sweep finally leads to a metal-insulator transition (MIT), whose onset was defined from the phase shift of the lock-in amplifier increasing above \SI{15}{\degree}. Sweeping $V\rm_{G}$ forward again results in a reversible decrease of $R\rm_{s}$ which overlaps with the previous backward sweep and the system is fully recovered when $V\rm_{G}$ is reapplied to \SI{50}{\volt}. We therefore classify $V\rm_{G}$ sweeps into three regimes, namely irreversible forward sweep (FS$\rm_{irrev}$), backward sweep (BS) and reversible forward sweep  (FS$\rm_{rev}$). $V\rm_{G}$ was then increased to \SI{200}{\volt} ($V\rm_{G}^{max2}$) to drive $E\rm_{F}$ well above the Lifshitz point. Similar reversible behavior is observed in BS$\rm_{2}$ and FS$\rm_{rev2}$. Back-gate tuning of MR in various regimes is shown in Fig. \ref{figmr}. Note for instance now the positive MR at \SI{50}{\volt} reverts to the single-band negative MR at \SI{10}{\volt} in the backward sweep regime BS$\rm_{1}$.

\begin{figure}[t]
	\centering
	\includegraphics[width=0.96\linewidth]{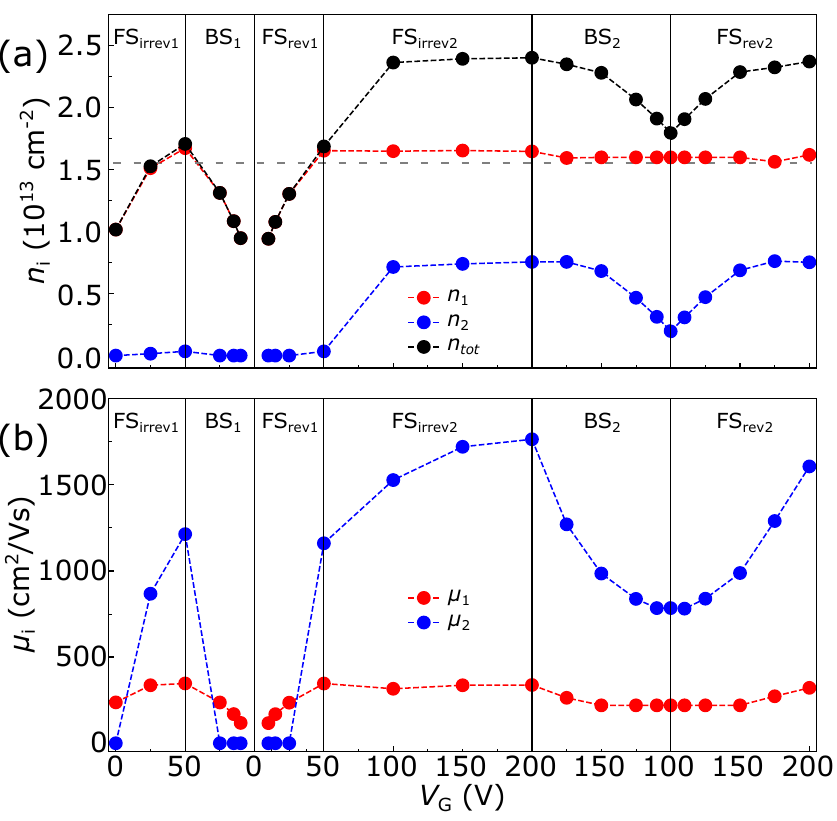}
	\caption{(a) $V\rm_{G}$ dependence of carrier densities. $n\rm_{1}$ and $n\rm_{2}$ stand for that of the low mobility carriers (LMC) and high mobility carriers (HMC), respectively. The total carrier density ($n\rm_{tot}$) is the sum of $n\rm_{1}$ and $n\rm_{2}$. The gray dash line represent the critical carrier density ($n\rm_{L}=$ \SI{1.51e13}{\centi\meter^{-2}}) for Lifshitz transition. (b) $V\rm_{G}$ dependence of mobilities, that of the LMC and HMC are labeled as $\mu\rm_{1}$ and $\mu\rm_{2}$, respectively.}
	\label{figNmu}
\end{figure}

\begin{figure}[t]
	\centering
	\includegraphics[width=0.96\linewidth]{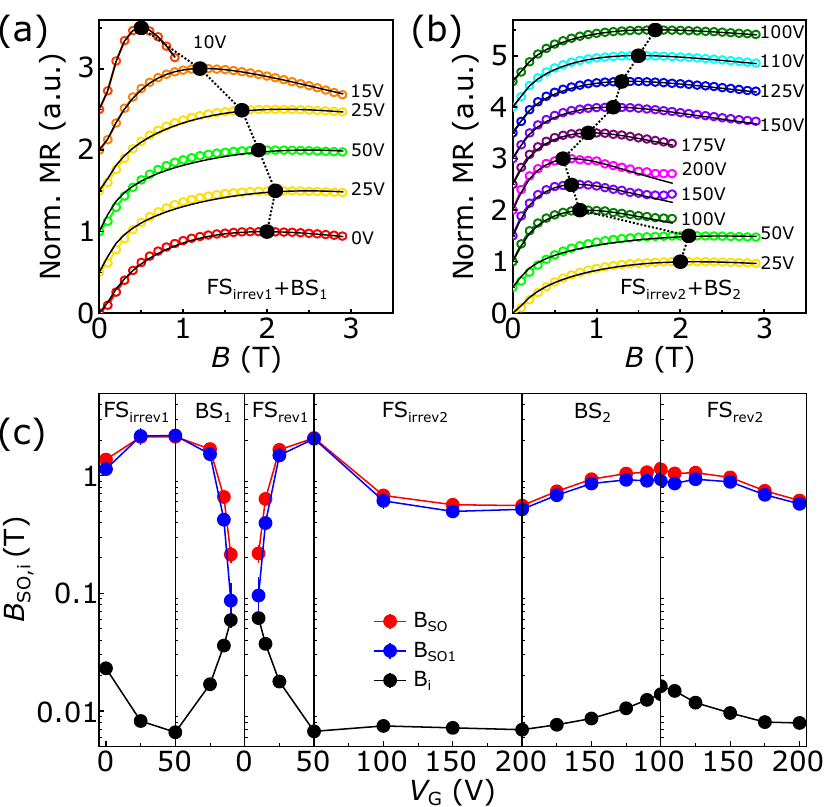}
	\caption{(a)-(b) Weak antilocalization (WAL) analysis in regimes (a) FS$\rm_{irrev1}$ and BS$\rm_{1}$ and (b) FS$\rm_{irrev2}$ and BS$\rm_{2}$. The solid circles correspond to experimental data and black lines to fits using the ILP model. The local maximum of each MR curve is plotted black. The MR curves are normalized to the local maxima and separated by an offset of 0.5. The black dashed line is a guide to the eye for the evolution of the local maxima. (c) Fitted characteristic magnetic fields as a function of $V\rm_{G}$. SO field $B{\rm_{SO}}$ is the sum of $B{\rm_{SO1}}$ (single spin winding) and $B{\rm_{SO3}}$ (triple spin winding). $B{\rm_{i}}$ is the inelastic scattering field.}
	\label{figfit}
\end{figure}

Fig. \ref{figNmu}(a) and \ref{figNmu}(b) present the $V\rm_{G}$ dependence of the carrier densities and mobilities. The values are extracted by fitting the magnetotransport data with a two-band model \cite{biscaras2012PRL,textsm}:
\begin{equation}
R_{xy}=\frac{B}{e}\frac{\frac{n_{1}\mu_{1}^{2}}{1+\mu_{1}^{2}B^{2}}+\frac{n_{2}\mu_{2}^{2}}{1+\mu_{2}^{2}B^{2}}}{(\frac{n_{1}\mu_{1}}{1+\mu_{1}^{2}B^{2}}+\frac{n_{2}\mu_{2}}{1+\mu_{2}^{2}B^{2}})^{2}+(\frac{n_{1}\mu_{1}^{2}B}{1+\mu_{1}^{2}B^{2}}+\frac{n_{2}\mu_{2}^{2}B}{1+\mu_{2}^{2}B^{2}})^{2}},
\end{equation}
with the constraint $1/R{\rm_{s}}(B=0)=en_{1}\mu_{1}+en_{2}\mu_{2}$, where $n\rm_{1}$ and $n\rm_{2}$ are the carrier densities of the low mobility carriers (LMC) and high mobility carriers (HMC), respectively, and $\mu_{1}$ and $\mu_{2}$ are the corresponding mobilities. For a reliable convergence $n\rm_{2}$ and $\mu_{2}$ are set to 0 in the one-band transport regime. The total carrier density ($n\rm_{tot}$) is the sum of $n\rm_{1}$ and $n\rm_{2}$. As shown in Fig. \ref{figNmu}(a), the critical carrier density ($n\rm_{L}$) corresponding to the Lifshitz transition is \SI{1.51e13}{\centi\meter^{-2}}, which is close to earlier reported results \cite{joshua2012NC}. The evolution of the carrier densities indicates that $E\rm_{F}$ approaches the Lifshitz point in regimes FS$\rm_{irrev1}$, FS$\rm_{rev1}$ and BS$\rm_{2}$ and departs from the Lifshitz point in regimes BS$\rm_{1}$, FS$\rm_{irrev2}$ and FS$\rm_{rev2}$. In Fig. \ref{figNmu}(b), it can be seen that $\mu_{1}$ almost stays unaffected above the Lifshitz transition whereas $\mu_{2}$ can be considerably changed by $V\rm_{G}$, reaching $\sim$\SI{1800}{\centi\meter^{2}/\volt\second} at \SI{200}{\volt}. It should be mentioned that there is a small upturn in $R\rm_{xy}$ which cannot be captured by the two-band model (for more details see Ref. \cite{textsm}). A similar feature has also been reported by other groups \cite{joshua2012NC,seiler2018PRB}, but its origin is still under debate. There are attempts to relate it to an unconventional anomalous Hall effect (AHE) \cite{ruhman2014PRB} or hole transport \cite{seiler2018PRB}, but we cannot get convincing fits using these models. In any case, we emphasize that the extraction of the parameters is not affected strongly by this feature.

In low-dimensional systems, the conductivity shows signatures of quantum interference between time-reversed closed-loop electron trajectory pairs. In the presence of SO coupling the pairs interfere destructively, leading to a positive MR in low magnetic field which is known as the WAL \cite{bergmann1984}. For a system with Rashba-type of SO coupling, the spin relaxation is described by the D'yakonov-Perel' (DP) mechanism \cite{dyakonov1971JETP}. The model for analyzing the WAL was established by Iordanskii, Lyanda-Geller and Pikus (ILP) \cite{iordanskii1994}. In this model, both the single and triple spin winding contributions at the Fermi surface have been taken into account. It should be noted that the ILP model is an effective single-band model, which means that above the Lifshitz point the fitted characteristic magnetic field for SO coupling is an effective field for both the $d_{xy}$ and $d_{xz,yz}$ bands. A model that considers multi-band effects is not available yet. The WAL correction to the magnetoconductivity is given by \cite{iordanskii1994,seiler2018PRB}:
\begin{equation}
\Delta\sigma(B) = -\frac{e^{2}}{2\pi h} [\mathcal{L}(B)-\mathcal{L}(0)+\psi(\frac{1}{2}+\frac{B{\rm_{i}}}{B})-\ln(\frac{B{\rm_{i}}}{B})],
\end{equation}
\begin{multline}
\mathcal{L}(B) = \frac{1}{a_{0}} + \frac{2a_{0}+1+\frac{B{\rm_{SO1}}+B{\rm_{SO3}}}{B}}{a_{1}(a_{0}+\frac{B{\rm_{SO1}}+B{\rm_{SO3}}}{B})-2\frac{B{\rm_{SO1}}}{B}} \\ + \sum_{n=1}^{\infty}\frac{3a_{n}^{2}+2a_{n}\frac{B{\rm_{SO1}}+B{\rm_{SO3}}}{B}-1-2(2n+1)\frac{B{\rm_{SO1}}}{B}}{(a_{n}+\frac{B{\rm_{SO1}}+B{\rm_{SO3}}}{B})a_{n-1}a_{n+1}-2\frac{B{\rm_{SO1}}}{B}[(2n+1)a_{n}-1]},
\end{multline}
where $\psi$ is the digamma function, $a_{n}=n+1/2+(B{\rm_{i}}+B{\rm_{SO1}}+B{\rm_{SO3}})/B$. The fitting parameters are the characteristic magnetic fields for the inelastic scattering $B{\rm_{i}}=\hbar/4eD\tau{\rm_{i}}$, and for the spin-orbit coupling $B{\rm_{SO}}_{n}=(\hbar/4eD)2\Omega_{n}^{2}\tau_{n}$ ($n$ = 1 or 3 for single or triple spin winding), where $D$ is diffusion constant, $\tau{\rm_{i}}$ and $\tau_{n}$ are relaxation times, and $\Omega_{n}$ is spin splitting coefficient. 

Fig. \ref{figfit}(a) and \ref{figfit}(b) depict WAL fits in the two FS$\rm_{irrev}$ and BS regimes. The solid black circles represent the local maxima of the MR curves. In principle, the SO coupling strength can be roughly estimated by the magnetic field ($B\rm_{max}$) where the local maximum appears \cite{liang2015PRB}. It can be clearly seen that $B\rm_{max}$ increases as $E\rm_{F}$ approaches the Lifshitz point (regimes FS$\rm_{irrev1}$ and BS$\rm_{2}$), while $B\rm_{max}$ decreases as $E\rm_{F}$ departs from the Lifshitz point (regimes BS$\rm_{1}$ and FS$\rm_{irrev2}$). The fitted values for the characteristic magnetic fields are plotted in Fig. \ref{figfit}(c), where $B{\rm_{SO}}$ is the sum of $B{\rm_{SO1}}$ and $B{\rm_{SO3}}$. In most cases $B{\rm_{SO3}}$ is much smaller than $B{\rm_{SO1}}$, indicating a single spin winding at the Fermi surface. The maximum SO coupling strength occurs near the Lifshitz point, agreeing with the evolution of $B\rm_{max}$. Driving $E\rm_{F}$ either above or below the Lifshitz point would lead to a decrease of the SO coupling strength. $B{\rm_{i}}$ increases when the carrier density is lowered, which is due to more accessible phonons contributing to the scattering process, and vice versa.  

If $B{\rm_{SO1}}$ is 0 and only $B{\rm_{SO3}}$ is present, the ILP formula could be reduced to a simpler model developed by Hikami, Larkin and Nagaoka (HLN) \cite{hikami1980}, in which the spin relaxation is described by the Elliot-Yafet (EY) mechanism \cite{elliott1954,yafet1963}. However, the HLN model yields inaccurate fits to our data, which is different from earlier reported results \cite{nakamura2012PRL,liang2015PRB}, where a triple spin winding has been found.

Our results manifest the nontrivial SO coupling mechanism at the LAO/STO interface predicted by theoretical works \cite{zhong2013PRB,khalsa2013PRB}. Applying an external electric field can tune the SO coupling, but its direct contribution is rather small. According to the Rashba theory for a free electron system, a typical electric field in experiments, \textit{e.~g.} \SI{100}{\volt}, only yields a spin splitting of $\sim$ \SI{e-8}{\milli\electronvolt} \cite{zhong2013PRB}, which is much smaller than the measured values that are of the order of meV \cite{caviglia2010PRL}. Instead the electric field-effect is indirect. It is the tuning of carrier densities and therefore band filling that significantly influence the SO coupling. 

In summary, we have performed magnetotransport experiments to study the Rashba SO couping effect in back-gated LAO/STO. By tuning the gate voltage, the Fermi energy has been driven to approach or depart from the Lifshitz point multiple times. We have done WAL analysis using the ILP model, which reveals a single spin winding at the Fermi surface. We have found that the maximum SO coupling occurs when the Fermi energy is near the Lifshitz point. Driving the Fermi energy above or below the Lifshitz point would result in a decrease of the coupling strength. Our findings provide valuable insights to the investigation and design of oxide-based spintronic devices.

We thank Thilo Kopp, Daniel Braak, Andrea Caviglia, Nicandro Bovenzi, Sander Smink, Aymen Ben Hamida and Prateek Kumar for useful discussions. This work is supported by the Netherlands Organisation for Scientific Research (NWO) through the DESCO program. We acknowledge the support of HFML-RU/NWO, member of the European Magnetic Field Laboratory (EMFL). P. S. is supported by the DFG through the TRR 80. C. Y. is supported by China Scholarship Council (CSC) with grant No. 201508110214.

\bibliographystyle{apsrev4-1}
\bibliography{LAO3}

\end{document}